\documentclass{llncs}
\usepackage{amssymb}
\usepackage{graphicx}
\usepackage{etex}
\usepackage{amsmath}
\usepackage{epsfig}
\usepackage{float}
\usepackage{floatflt}
\usepackage{booktabs}
\usepackage{paralist}
\usepackage{pgfplots}
\usepackage{tikz}
\usepackage{setspace}
\usepackage{comment}
\usepackage{multirow}
\usepackage{caption}
\usepackage[lofdepth,lotdepth]{subfig}
\usepackage[export]{adjustbox}
\usepackage[font=small,labelfont=bf]{caption}
\usepackage{etex}
\usepackage{tabu}
\usepackage{color,soul}
\usepackage{relsize}
\usepackage{todonotes}
\usepackage{hyperref}

\newcolumntype{H}{>{\setbox0=\hbox\bgroup}c<{\egroup}@{}}

\usepackage{url}

\makeatletter

\begin{document}

\title{Studying Catastrophic Forgetting  in Neural Ranking Models}
%
%

\newcommand{\emailsize}{\fontsize{10pt}\times}

\author{Jesús Lovón-Melgarejo$^{1}$ ~~ Laure Soulier$^{2}$ ~~
Karen Pinel-Sauvagnat$^{1}$ ~~
Lynda Tamine$^{1}$ \\
\mbox{}\\
$^1$Université Paul Sabatier, IRIT, Toulouse, France \\
$^2$Sorbonne Université, CNRS, LIP6, F-75005 Paris, France, \\
\emailsize{\texttt{\{jesus.lovon, sauvagnat, tamine\}@irit.fr, laure.soulier@lip6.fr}}}
\institute{\mbox{}}

\maketitle

\begin{abstract}
Several deep neural ranking models have been proposed in the recent IR literature. While their transferability to one target domain held by a dataset has been widely addressed using traditional domain adaptation strategies, the question of their cross-domain transferability is still under-studied. We  study here  in what extent neural ranking models catastrophically forget old knowledge acquired from previously observed domains after acquiring new knowledge, leading to performance decrease on those domains. Our experiments show that the effectiveness of neural IR ranking models is achieved at the cost of catastrophic forgetting and that a lifelong learning strategy using a cross-domain regularizer successfully mitigates the problem. Using an explanatory  approach built on a regression model, we also show the effect of domain characteristics on the rise of catastrophic forgetting. We believe that the obtained results can be useful for both theoretical and practical future work in neural IR.   
\end{abstract}

\vspace{-0.5cm}
\begin{sloppypar}
\section{Introduction }
\vspace{-0.3cm}


Neural ranking models have been increasingly adopted in the information retrieval (IR) and natural language processing (NLP) communities for a wide range of data and tasks \cite{mitra2018,Onal2017NeuralIR}. One common underlying issue is that  they learn relationships that may hold only in the domain from which the training data is sampled, and generalize poorly in unobserved domains\footnote{According to Jialin and Qiang \cite{Jialin10},  a domain consists of at most two components: a feature space  over a dataset and a marginal probability distribution within a task. } \cite{2018Chen,Onal2017NeuralIR}. 
To enhance the transferability of neural ranking models from a source domain to a target domain, transfer learning strategies  such as fine-tuning  \cite{1904-06652}, multi-tasking \cite{LiuGHDDW15}, domain adaptation \cite{Jialin10}, and more recently adversarial learning \cite{cohen2018}, have been widely used\footnote{We consider the definition of transfer learning in \cite{Jialin10} (Figure 2). Please note that several other  definitions exist \cite{gulrajani2021in}.}.  However, these strategies have by essence two critical limitations reported in the machine learning literature  \cite{2018Chen,KirkpatrickPRVD16}. The first one, as can be acknowledged in the NLP and IR communities \cite{cohen2018,LiuGHDDW15},   is that they require all the domains to be  available simultaneously at the learning stage (except the fine-tuning).  The second limitation, under-studied in both communities, is that the model leans to \textit{catastrophically forget} existing knowledge (source domain) when the learning is transferred to new knowledge (target domain) leading to a significant drop of performance on the source domain. These limitations are particularly thorny when considering  open-domain IR tasks including, but not limited to, conversational search.
In the underlying settings (e.g., QA  systems and chatbots \cite{hancock-etal-2019-learning,LiMCRW17a,abs-1802-06024,roller2020open}),  neural ranking models are expected to continually learn features from online information streams, sampled from either observed or unobserved domains, and to scale across different domains but without forgetting previously learned knowledge.  

\textit{Catastrophic forgetting} is a long-standing problem addressed in machine learning    using \textit{lifelong learning}  approaches \cite{2018Chen,ParisiKPKW19}. It has been particularly   studied in  neural-network based classification tasks in computer vision \cite{KirkpatrickPRVD16,8107520} and more recently in  NLP \cite{journals/corr/abs-1906-01076,2020arXiv200604884M,thompson-etal-2019-overcoming,wiese-etal-2017-neural-domain}. However, while previous work 
showed that the level of catastrophic forgetting is significantly impacted by dataset features and network architectures,  we are not aware of any existing research in IR providing  clear lessons about the transferability of neural ranking models across domains, nor basically showing if state-of-the-art neural ranking models  are  actually faced with the catastrophic forgetting problem and how to overcome it if any. Understanding the conditions under which these models forget accumulated knowledge and whether a lifelong learning strategy is a feasible way for improving their effectiveness, would bring important lessons  for both practical and theoretical work in IR. This work contributes to fill this gap identified  in the neural IR literature, by studying the transferability of ranking models. We put the focus on  catastrophic forgetting which is the bottleneck of lifelong learning. 
 
 The main contributions of this paper are as follows. 1) We  show the occurrence of catastrophic forgetting in neural ranking models.   We investigate the transfer learning of  five representative  neural ranking models  (DRMM\cite{Guoetal16},  PACRR\cite{hui2017pacrr},  KNRM\cite{xiong2017end},  V-BERT\cite{macavaney2019cedr}  and CEDR \cite{macavaney2019cedr}) over streams of datasets from different domains\footnote{In our work, different domains refer to different datasets characterized by different data distributions w.r.t. to their source and content  as defined in \cite{Jialin10}.}  (MS MARCO \cite{bajaj2016ms}, TREC Microblog \cite{trecML12} and TREC COVID19 \cite{wang2020cord}); 2) We identify domain characteristics such as relevance density  as signals of catastrophic forgetting  ; 3) We show  the effectiveness of constraining the objective function of the neural IR models  with a forget cost term, to mitigate the catastrophic forgetting. 
 
\vspace{-0.4cm}
\section{Background and Related Work }
\vspace{-0.1cm}

\subsubsection{From Domain Adaptation to Lifelong Learning of Neural Networks.} 
\vspace{-0.1cm}
Neural networks are learning systems that must commonly, on the one hand, exhibit the ability to acquire new knowledge and, on the other hand, exhibit robustness by refining knowledge while maintaining stable performance on existing knowledge. While the acquisition of new knowledge gives rise to the well-known \textit{domain shift} problem \cite{Quionero09}, maintaining model performance on existing knowledge is faced with the \textit{catastrophic forgetting} problem. Those problems have been respectively tackled using \textit{domain adaptation} \cite{Jialin10}  and \textit{ lifelong learning} strategies \cite{2018Chen,ParisiKPKW19}.  Domain  adaptation,  commonly known as a specific setting of  \textit{transfer learning} \cite{Jialin10}, includes machine learning methods (e.g., fine-tuning    \cite{wiese-etal-2017-neural-domain} and multi-tasking \cite{LiuGHDDW15})  that  assume that the source and the  target domains from which are sampled respectively the training and testing data might have different distributions. 
 By applying a transfer learning method, a neural model 
 should acquire new specialized knowledge from the target domain leading to optimal performance on it. \\ 
 One of the  main issues behind common transfer learning approaches is catastrophic forgetting  \cite{FRENCH1999128,Goodfellow2014AnEI}:  the newly acquired knowledge interfers with, at the worst case, overwrites, the existing knowledge leading to performance decrease on information sampled from the latter.
Lifelong learning \cite{2018Chen,ParisiKPKW19} tackles this issue by enhancing the models with the ability to continuously learn over time and accumulate knowledge from streams of information sampled across domains, either previously observed or not. 
The three common lifelong learning approaches  are \cite{ParisiKPKW19}: 1) regularization that  constrains the objective function with a forget cost term \cite{KirkpatrickPRVD16,8107520,wiese-etal-2017-neural-domain}; 2) network expansion that adapts the network architecture to new tasks by adding neurons and layers  \cite{cai-etal-2019-adaptive,RusuRDSKKPH16}; and 3) memory models that retrain the network using instances selected from a memory drawn from different data distributions \cite{Ashar20,journals/corr/abs-1906-01076}. 
\vspace{-0.4cm}
\subsubsection{On the Transferability of Neural Networks in NLP and IR.}   
Transferability of  neural networks  has been particularly studied in classification tasks,  first in computer vision  \cite{Bengio11,Yosinski14} and then only recently in NLP \cite{jha2020does,mosbach2020stability,mou-etal-2016-transferable}. 
For instance, Mou et al. \cite{mou-etal-2016-transferable} investigated the transferability of neural networks in sentence classification and sentence-pair classification  tasks. 
One of their main findings is that transferability across domains  depends on the level of similarity between the considered tasks.  Unlikely, previous work in IR which mainly involves ranking tasks,   have only casually applied transfer learning methods (e.g.,  fine-tuning \cite{1904-06652}, multi-tasking \cite{LiuGHDDW15}  and  adversarial learning \cite{cohen2018}) without bringing generalizable lessons about the transferability of neural ranking models. One consensual result reported across previous research in the area, is that traditional retrieval models (e.g., learning-to-rank models \cite{Liu09})   that make fewer distributional assumptions, exhibit more robust cross-domain performances \cite{cohen2018,Onal2017NeuralIR}. Besides, it has been shown that the ability of neural ranking models to learn new features may be achieved at the cost of poor performances on domains not observed during training  \cite{mitra2018}. Another consensual result  is that although  embeddings are trained using large scale corpora, they are generally sub-optimal for domain-specific ranking tasks \cite{Onal2017NeuralIR}. \\
Beyond domain adaptation, there is a recent research trend  in NLP toward lifelong learning of neural networks, particularly in machine translation  \cite{thompson-etal-2019-overcoming}, and language understanding tasks \cite{2020arXiv200604884M,wiese-etal-2017-neural-domain,ijcai2018-627}. 
For instance, Xu et al. \cite{ijcai2018-627}  recently revisited the domain transferability of traditional word embeddings \cite{Mikolov13} and proposed \textit{lifelong domain embeddings} using a meta-learning approach. The proposed meta-learner  is fine-tuned to identify similar contexts of the same word in both past domains and the new observed domain. Thus, its
inference model is able to compute the similarity scores on pairs of feature vectors representing the same word across domains.   These embeddings have been successfully applied to a topic-classification task. 
Unlikely, catastrophic forgetting and lifelong learning are still under-studied in IR. We believe that a thorough analysis of the transferability of neural ranking models from a lifelong learning perspective would be desirable for a wide range of emerging open-domain IR applications including but not limited to conversational search \cite{hancock-etal-2019-learning,abs-1802-06024,LiMCRW17a,roller2020open}. 
\vspace{-0.4 cm}
\section{Study Design }
\vspace{-0.3cm}
Our study mainly addresses  the following research questions:\\
\textbf{RQ1: }Does catastrophic forgetting occur in neural ranking models?\\
\textbf{RQ2}: What are the dataset characteristics that predict catastrophic forgetting?  \\
\textbf{RQ3:} Is a regularization-based lifelong learning method effective to mitigate catastrophic forgetting in neural ranking models?
\vspace{-0.3cm}
\subsection{Experimental Set Up}
\vspace{-0.2cm}
Given a neural model $M$ designed for an ad-hoc ranking task,  the primary objectives of our experiments are twofold: O1) measuring the catastrophic forgetting of model $M$ while applying a domain adaptation method $\mathcal{D}$,  in line of RQ1 and RQ2; and O2) evaluating the effect of a lifelong learning  method $\mathcal{L}$  to mitigate catastrophic forgetting in model $M$, in line of RQ3.   We assume that model  $M$  learns a  ranking task across a stream of $n$ domain datasets $\{D_1, \dots , D_n\}$ coming in a sequential manner one by one. At a high level, our experimental set up is:

\vspace{2px}
\noindent
\fbox{
\begin{minipage}{1\textwidth}
\footnotesize
\begin{enumerate}
 \item Set up an ordered dataset stream \textbf{setting}  $D_1\rightarrow \dots    D_{n-1}\rightarrow D_n$ 
 \item Learn oracle models $M_i^*, i=1\dots n $, with parameters $\hat{\theta}^{i*}  $   by training \textbf{the neural ranking model  $M$} on training instances of \textbf{dataset $D_i, i=1 \dots n$}.
 \item Measure the retrieval performance $R_{i,i}^*$ of each oracle model $M_i^*$ on testing instances of the same dataset  $D_i$.
 \item Launch a \textbf{domain adaptation  method $\mathcal{D}$}  w.r.t. to objective O1  (resp. a \textbf{lifelong learning method $\mathcal{L}$} w.r.t. to objective O2) along the considered setting  as follows:
 \begin{itemize}
 \item Initialize ($k=1$) model $M_k$,  with  $\hat{\theta}^{1*}$,  parameters of model $M_1^*$ (trained on the dataset base $D_{1}$).
  \item Repeat  
  \begin{itemize}
 \item Apply to model $M_k$  a method $\mathcal{D}$ w.r.t to objective O1 (resp.  method $\mathcal{L}$ w.r.t. to objective O2) to transfer knowledge to the right dataset $D_{k+1}$ (forward transfer). The resulting model  is noted $M_{k+1}$ with parameters $\hat{\theta}^{k+1}$. Its performance on dataset $D_{k+1}$ is noted  $R_{k+1,k+1}$.
 \item Measure the retrieval performance $R_{k+1,k}$ of model $M_{k+1}$ obtained  on the testing instances of left dataset  $D_{k}$ (backward transfer) 
 \item Move to the next right dataset :  $k=k+1$
 \end{itemize}
  \item Until the end of the dataset stream setting ($k=n$).
  \end{itemize}
  \item \textbf{Measure catastrophic forgetting} in model $M$. 
\end{enumerate}
\end{minipage}
}
\normalsize

\begin{figure}[!t]
    \centering
    \includegraphics[width=0.80\linewidth]{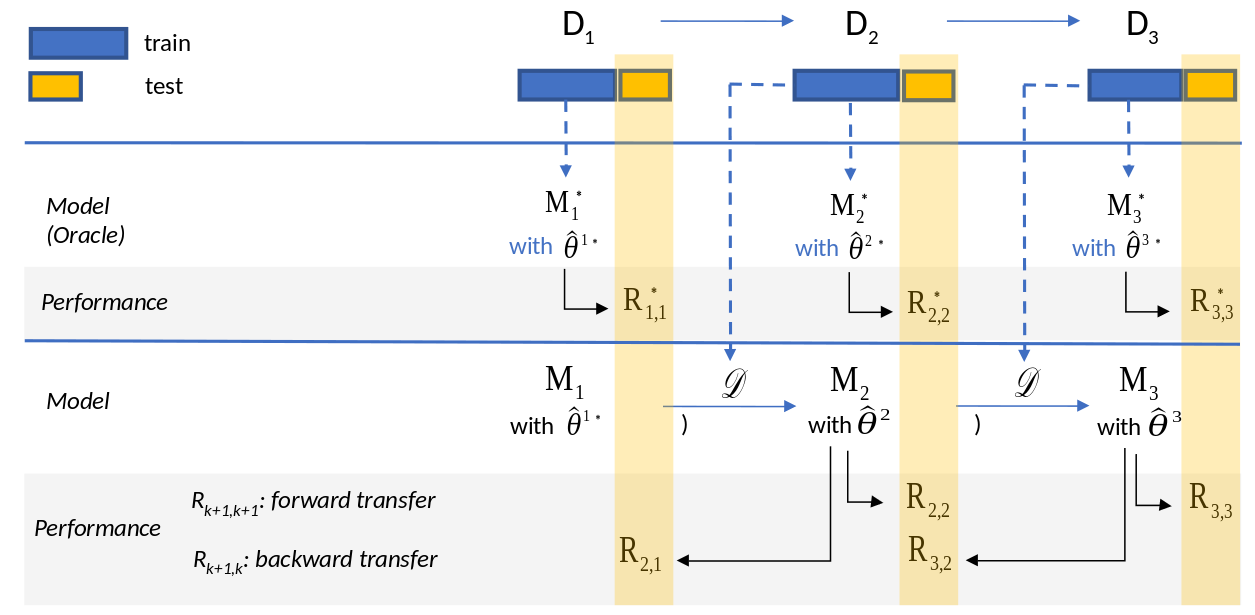}
    \vspace{-0.2cm}
    \caption{Experimental pipeline using a 3-dataset stream setting 
    for a given model M}
    \vspace{-0.4cm}
    \label{fig:pipeline}
\end{figure}

\vspace{0.5cm}
This experimental pipeline, illustrated in Figure \ref{fig:pipeline}, follows general guidelines adopted in previous work \cite{Ashar20,KemkerMAHK18,8107520}.  We detail below the main underlying components highlighted in bold. 

\paragraph{\textbf{Neural ranking models}} We evaluate catastrophic forgetting in  five  (5) state-of-the-art models selected from a list of models criticallly evaluated in Yang et al. \cite{yang2019critically}: 1) interaction-based models: DRMM \cite{Guoetal16} and PACRR \cite{hui2017pacrr} and KNRM \cite{xiong2017end};  2) BERT-based models: Vanilla BERT \cite{macavaney2019cedr} and CEDR-KNRM \cite{macavaney2019cedr}.  We use the OpenNIR framework \cite{macavaney:wsdm2020-onir} that provides a complete neural ad-hoc document ranking pipeline. Note that in this framework, the neural models are trained by linearly combining their own score ($S_{NN}$) with a BM25 score ($S_{BM25}$).
\vspace{-0.2cm}
\paragraph{\textbf{Datasets and settings.}} We use the three  following datasets: 1) MS MARCO (\textit{\textbf{ms}}) \cite{bajaj2016ms} a passage ranking dataset which includes more than 864 K question-alike queries sampled from the Bing search log and a large-scale web document set including 8841823 documents; 2) TREC Microblog (\textit{\textbf{mb}}) \cite{Microblogoverview13}, a real-time ad-hoc search dataset from TREC Microblog 2013 and 2014, which contains a public Twitter sample stream between February 1 and March 31, 2013 including 124969835 tweets and 115 queries submitted at a specific point in time; 3) TREC CORD19 (\textit{\textbf{c19}}) \cite{wang2020cord} an ad-hoc document search dataset which contains 50 question-alike queries and a corpora of 191175 published research articles dealing with SARS-CoV-2 or COVID-19 topics. It is worth mentioning that these datasets fit with the requirement of cross-domain adaptation \cite{Jialin10} since they have 
significant differences in both their content and sources.  
Besides, 
we consider four settings (See Table \ref{tab:omega_results_map}, column \textbf{"Setting"}) among which three 2-dataset ($n=2$) and one 3-dataset ($n=3$) settings. As done in previous work \cite{Ashar20,8107520}, these settings  follow the patterns ($D_1\rightarrow D_2$) or ($D_1\rightarrow D_2 \rightarrow D_3$) where dataset orders are based on the decreasing sizes of the training sets assuming that larger datasets allow starting with well-trained networks.
\vspace{-0.2cm}
\paragraph{\textbf{ Domain adaptation and lifelong learning methods. }}
We adopt fine-tuning (training on one domain and fine-tuning on the other) as the representative domain adaptation task  $\mathcal{D}$ since it suffers from the catastrophic forgetting problem \cite{Ashar20,KirkpatrickPRVD16}. Additionally, we adopt the Elastic Weight Consolidation (EWC) \cite{KirkpatrickPRVD16} as the lifelong learning method $\mathcal{L}$.   The EWC constrains the loss function with an additional forget cost term that we add to the objective function of each of the five neural models studied in this work.     Basically speaking, EWC constrains the neural network-based model to remember knowledge acquired on left datasets by reducing the overwriting of its most important parameters  as:
\vspace{-0.2cm}
\begin{equation}
    \mathcal{L}(\hat{\theta}^{k}) = \mathcal{L}(\hat{\theta}^{k}) + \Sigma_{1\leq i<k} \frac{\lambda}{2} \mathcal{F}_i(\hat{\theta}^{k} - \hat{\theta}^{i})^2
    \vspace{-0.3cm}
\end{equation}
where  $\mathcal{L}(\hat{\theta}^{k})$ is the loss of the neural ranking model with parameters $\theta^{k}$ obtained right after learning on ($D_k$), $\lambda$ is the importance weight of the models parameters trained on left datasets   ($D_i, i<k$)  with the current one ($D_k$), $\mathcal{F}$ is the Fisher information matrix. 

\vspace{-0.2cm}

\paragraph{\textbf{ Measures.} } Given the setting ($D_1\rightarrow \dots \rightarrow D_n$), we use  the \textit{remembering} measure (REM) derived from the \textit{backward transfer measure} (BWT) proposed by Rodriguez et al. \cite{Diaz18} as follows:

$\bullet$ \textbf{BWT}: measures the intrinsic effect (either positive or negative) that learning a model $M$ on a new dataset (right in the setting) has on the model performance obtained on an old dataset (left in the setting), referred as \textit{backward transfer}. Practically, in line with a lifelong learning perspective,  this measure averages along the setting the differences between the performances of the model obtained right after learning on the left dataset  and the  performances of the oracle model trained and tested on the same left dataset. Thus, while positive values  handle positive backward transfer, negative values handle catastrophic forgetting.  Formally, the BWT measure is computed as:
 \begin{equation}
   BWT=  \frac{\sum_ {i=2}^n\sum_ {j=1}^{i-1}(R_ {i,j}-R_ {j,j}^* )}{\frac{n(n-1)}{2}} 
\end{equation}
$R_ {i,j}$ is the performance measure of model $M_i$ obtained right after learning on on dataset $D_j$.   $R_ {j,j}^*$ is the performance of the oracle model $M_j^*$ trained on dataset $D_j$ and tested on the same dataset. To make fair comparisons between the different studied neural models, we normalize the differences in performance ($R_ {i,j}-R_ {j,j}^* $)  on model agnostic performances obtained using  $BM25$ model on each left dataset $D_j$.  In our work, we use the standard IR performance measures MAP, NDCG@20 and P@20 to measure $R_ {i,j}$  but we only report the REM values computed using the MAP measure, as they all follow the same general trends.   

\indent $\bullet$ \textbf{REM}: because the BWT measure assumes either positive values for positive backward transfer and negative values for catastrophic forgetting, it allows to map with a positive remembering value in the range $[0, 1]$ as follows:
  \begin{equation}
   REM=  1-  \lvert min(BWT,0)\rvert 
\end{equation}
A REM value equals to 1 means that the   model does not catastrophically forget. 

To better measure the intrinsic ability of the neural ranking models to remember previously acquired knowledge, we deploy in the OpenNIR framework two runs for each neural model based on the score combination 
($score_G = \alpha \times  S_{NN}+ (1-\alpha) \times S_{BM25}$). The first one by considering the neural model after a re-ranking setup ($0<\alpha<1$) leading to compute an overall $REM$ measure on the ranking model. The second one by only considering the neural ranking based on the $S_{NN}$ score by totally disregarding the BM25 scores ($\alpha=1$). $REMN$ denotes the remembering measure computed in this second run.

%

\vspace{-0.3cm}
\subsection{Implementation details}
\vspace{-0.2cm}
 We use the OpenNIR framework with default parameters and the pairwise hinge loss function \cite{dehghani2017neural}. To feed the neural ranking models, we use the GloVe pre-trained embeddings (42b tokens and 300d vectors). The datasets are split into training and testing instance sets. For MS MARCO, we use the default splits provided in the dataset. For TREC CORD19 and TREC Microblog, where no training instances are provided, we adopt the splits by proportions leading to 27/18 and 92/23 training/testing queries respectively.  In practice, we pre-rank documents using the BM25 model. For each  relevant document-query  pair (positive pair), we randomly sample a document for the same query with a lower relevance score to build the negative pair. We re-rank the top-100 BM25 results and use $P@20$ to select the best-performing model. 
For each dataset, we use the optimal BM25 hyperparameters selected using grid-search. In the training phase, we consider a maximum of 100 epochs or early-stopping  if no further improvement is found. Each epoch consists of 32 batches of 16 training pairs. All the models are optimized using Adam \cite{kingma2014adam} with a learning rate of $0.001$. BERT layers are trained at a rate of $2\mathrm{e}{-5}$ following previous work \cite{macavaney2019cedr}.  For the EWC, we fixed $\lambda=0.5$. The code is available at \url{https://github.com/jeslev/OpenNIR-Lifelong}.


\vspace{-0.3cm}
\section{Results}
\vspace{-0.3cm}
\subsection{Empirical Analysis Of Catastrophic Forgetting in Neural Ranking Models}

 \begin{table}[!t]
 \centering
 \small
     \caption{Per model-setting results in our fine-tuning  and EWC-based lifelong learning experiments. All the measures are based on the MAP@100 metric. The improvements $\Delta_{MAP(MAPN)}$ and $\Delta_{REM(REMN)}$ are reported in percent   (\%).}
          \begin{adjustbox}{max width=\textwidth}

      \begin{tabular}{ p{11mm} p{24mm}| p{20mm} H  p{22mm}p{8mm}| p{20mm}  p{20mm}  p{8mm} } 
     \toprule
      \textbf{Model}  & \textbf{Setting} & \multicolumn{4}{c|}{\textbf{Fine-tuning}} & \multicolumn{3}{c}{\textbf{EWC-based lifelong learning}}   \\
      & &  \textit{REM(REMN)}&   \textit{$\Delta_{MAP(MAPN)}$} &\textit{$\Delta_{MAP(MAPN)}$}  & \textit{PR}& \textit{REM(REMN)} &   \textit{$\Delta_{REM(REMN)}$} 
      & \textit{PR}  \\ 
     \midrule
\multirow{5}{*}{DRMM} & $ms\rightarrow c19$   & 1.000(1.000) & 0.023(-0.715)&+2.2(-73.6)&1.008  &1.000(1.000)      &0(0)  &1.005\\
& $ms\rightarrow mb$                      & 0.962(0.943)   & -0.017
(-0.793) &+2.2(-73.6)&1.021  &0.971(0.974)  &+0.9(\textbf{+3.3})   &1.011\\
& $mb\rightarrow c19$                     & 1.000(0.965)  &-0.017
(-0.112)&-1.7(-7.7) &0.993  &1.000(0.662)   &0(-31.4)  &0.995\\
& $ms\rightarrow mb\rightarrow c19$                  & 0.976(0.938)  &-0.008(-0.726)&+2(-73.6) &1.011  &0.979(1.000)  &\textbf{+0.3}(\textbf{+6.6})  &1.004\\
\midrule
\multirow{5}{*}{PACRR} & $ms\rightarrow c19$  & 1.000(0.760)  &0.026(-0.54) &+2.5(-30.1)&1.000  &1.000(0.756)  &0(-0.5)  &1.000\\
& $ms\rightarrow mb$                      & 1.000(1.000)  &0.026(-0.243)& +2.5(-30.1)&0.999 &1.000(1.000)     &0(0)  &1.014\\
& $mb\rightarrow c19$                    & 1.000(0.523)  &-0 (-37.6) &0(+10)&1.000  &1.000(0.940)   &0(\textbf{+79.7})  &1.002\\
& $ms\rightarrow mb\rightarrow c19$                   & 1.000(0.759) & 0.026(-0.636) &+2.5(-30)&1.000  &1.000(0.874)   &0(\textbf{+15.2})  &1.015\\
\midrule
\multirow{5}{*}{KNRM} & $ms\rightarrow c19$   & 1.000(1.000)  &-0.032
(-0.862)&-12.1(-89)&1.069  &1.000(1.000)     &0(0)  &1.058\\
&$ms\rightarrow mb$                     & 1.000(1.000)   &-0.088
(-0.784) &-12.1(-89)&0.991  &1.000(1.000)     &0(0)   &0.991\\
& $mb\rightarrow c19$                   & 1.000(0.810)  &0.011
(-0.328) &-2(-13.8)&1.135  &1.000(0.902)   &0(\textbf{+11.4})  &1.141\\
& $ms\rightarrow mb\rightarrow c19$                  & 1.000(1.000) & -0.045(-0.802) &-12.1(-89)&1.086  &1.000(0.963)  &0(-3.7)  &1.087\\
\midrule
\multirow{5}{*}{VBERT} & $ms\rightarrow c19$  & 0.930(1.000)   & -0.175(0.006) &-10.6(0)&1.028  &1.000(1.000)    &\textbf{+7.5}(0)   &0.990\\
& $ms\rightarrow mb$                    & 1.000(0.883)   &-0.003
(-0.111) &-10.6(0)&1.030  &1.000(1.000)    &0(\textbf{+13.3})   &0.992\\
& $mb\rightarrow c19$                    & 0.913(1.000)  &0.086(0.258)&+17.4(+25.8)&0.963  &1.000(1.000)    &\textbf{+9.5}(0)  &1.010\\
& $ms\rightarrow mb\rightarrow c19$                   & 0.989(0.922)   &-0.145(-0.111)&-10.6(0)&1.011  &1.000(1.000)  &\textbf{+1.1}(\textbf{+8.5})  &0.987\\
\midrule
\multirow{5}{*}{CEDR} & $ms\rightarrow c19$   & 0.826(1.000)   & -0.148(0.142) &+2.6(+14.2)&1.013 &1.000(1.000)    &\textbf{+21.1}(0)  &1.008\\
& $ms\rightarrow mb$                     & 0.510(0.920)   &-0.463
(0.062) &+2.6(+14.2)&1.003  &1.000(1.000)  &\textbf{+96.1}(\textbf{+8.7})   &0.976\\
& $mb\rightarrow c19$                   & 0.940(1.000)  &0.136(
0.292) &+19.6(+29.2)&1.011  &1.000(1.000)    &\textbf{+6.4}(0)  &0.984\\
& $ms\rightarrow mb\rightarrow c19$                  & 0.771(0.946)   &-0.194(0.062)&+2.6(+14.2)&0.996  &0.891(1.000)  &\textbf{+15.6}(\textbf{+5.7})  &0.961\\
\bottomrule
     \label{tab:omega_results_map}
     \end{tabular}
     \vspace{-1.2cm}
     \end{adjustbox}
     \vspace{-1.2cm}
 \end{table}

\vspace{-0.2cm}
\subsubsection{Within- and Across-Model Analysis  (RQ1).}
Our objective here is to investigate whether each of the studied neural models suffer from catastrophic forgetting while it is fine-tuned over a setting  ($D_1 \rightarrow D_2$ or $D_1 \rightarrow D_2 \rightarrow D_3$).
To carry out a thorough analysis of each model-setting pair, we compute the following measures in addition to the REM/REMN measures: 1) the MAP@100 performance ratio ($PR=  \frac{1}{(n-1)}\sum_ {i=2}^n \frac{R_ {i,i}}{R_ {i,i}^* }$) of the model learned on the right dataset and normalized on the oracle model performance;
2) the relative improvement in MAP@100 $\Delta_{MAP}$ (resp. $\Delta_{MAPN}$) achieved with the ranking based on the global relevance score $Score_G$ (resp. $Score_{NN} $)  trained and tested on the left dataset over the performance of the BM25 ranking obtained on the same testing dataset. 
Table \ref{tab:omega_results_map} reports all the metric values 
for each model/setting pairwise. In line with this experiment's objective, we focus  on the \textbf{"Fine-tuning"} columns. 

Looking first at the $PR$ measure reported in Table \ref{tab:omega_results_map}, we notice that  it is greater than $0.96$ in $100\%$ of the settings, showing that the fine-tuned models are successful on the right dataset, and thus allow a reliable investigation of catastrophic forgetting as outlined in previous work \cite{mosbach2020stability}.   
It is worth recalling that the general evaluation framework is based on a pre-ranking (using the BM25 model) which is expected to provide positive training instances from the left dataset to the neural ranking model being fine-tuned on the right dataset. 
The joint comparison of the $REM$ (resp. $REMN$) and $\Delta_{MAP}$ (resp.$\Delta_{MAPN}$ ) measures lead us to highlight the following statements:

$\bullet$ We observe that only CEDR and VBERT models achieve positive improvements w.r.t to both the global ranking ($\Delta_{MAP}$ : $+19.6\%$, $+17.4\%$ resp.)  and the neural ranking ($\Delta_{MAP}$: $+29.2\%$, $+25.8\%$ resp.),  particularly under the setting where $mb$ is the left dataset ($mb\rightarrow c19$).  
Both models are able to bring effectiveness gains additively to those brought by the exact-based matching signals in BM25. 
These effectiveness gains can be viewed as new  knowledge 
in the form of semantic matching signals which are successfully transferred to the left dataset ($c19$) while maintaining stable performances on the left dataset ($mb$)  (REMN=0.940 and 0.913 for resp. CEDR and VBERT).    This result is consistent with previous work suggesting that the regularization used in transformer-based models  has an effect of alleviating catastrophic forgetting  \cite{LeeCK20}.

 $\bullet$ We notice that the CEDR model achieves positive improvements w.r.t to  the neural ranking score ($\Delta_{MAPN}$: $+14.2\%$) in all the settings (3/4) where $ms$ is the left dataset while very low improvements are achieved  w.r.t. to the global score ($\Delta_{MAP}$: $+2.6\%$). We make the same observation for the PACRR model but only for 1/4 model-setting pair ($\Delta_{MAPN}$: $+10\%$ vs. $\Delta_{MAPN}$: $0\%$) with $mb$ as the left dataset. 
 Under these settings, we can see that even the exact-matching signals brought by the BM25 model are very moderate (leading to a very few positive training instances), the CEDR and, to a lower extent, the PACRR models, are able to inherently bring significant new knowledge in terms of semantic matching signals at however the cost  of significant forget on the global ranking for CEDR (REM is the range $[0.510;0.826]$) and on the neural ranking for PACRR (REM=0.523). 
 
$\bullet$ All the models (DRMM, PACRR, KNRM and VBERT (for 3/4 settings)  that do not significantly beat the BM25 baseline either by using the global score ($\Delta_{MAP}$ in the range $[-12.1\%;+2.2\%]$)  nor  by  using the neural score ($\Delta_{MAPN}$ in the range $[-89\%;+0\%]$), achieve near upper bound of remembering (both REM and REMN are in the range $[0.94;1]$). Paradoxically, this result does not allow us to argue about the ability of these models to retain old knowledge. Indeed, the lack or even the low improvements over both the exact matching (using the BM25 model) and the semantic-matching  (using the neural model) indicate that a moderate amount of new knowledge or even no knowledge about effective relevance ranking has been acquired from the left dataset. Thus, the ranking performance  of the fine-tuned model on the left dataset only depends on the level of mismatch between the data available in the right dataset for training and the test data in the left dataset. We can interestingly see that upper bound remembering performance  ($REM=1$) is particularly achieved when $ms$ is the left dataset (settings $ms\rightarrow c19$, $ms\rightarrow mb$, $ms\rightarrow mb\rightarrow  c19$). This could be explained by the fact that the relevance matching signals learned  by the neural model in in-domain knowledge do not degrade its performances on general-domain knowledge. 



Assuming  a well-established practice in neural IR which consists in linearly interpolating  the neural scores with the exact-based matching scores (e.g., BM25 scores),  these observations give rise to three main findings: 1)  the more  a neural ranking model  is inherently effective in learning  additional semantic matching signals, the more likely it catastrophically forgets. In other terms, intrinsic effectiveness of neural ranking models comes at the cost of forget;  2) transformer-based language models such as CEDR and VBERT exhibit a good balance between effectiveness and  forget as reported in previous work in NLP \cite{mosbach2020stability}; 3) given the variation observed in REM and REMN, there is no clear trend about which ranking (BM25-based ranking vs. neural ranking) impacts more importantly the level of overall catastrophic forgetting of the neural models

\vspace{-0.5cm}

\subsubsection{ Across Dataset Analysis (RQ2).}
Our objective here is to identify catastrophic forgetting signals from the perspective of the left dataset. As argued in \cite{Ado12}, understanding the relationships between data characteristics and catastrophic forgetting allows  to anticipating the choice of datasets in lifelong learning settings regardless of the neural ranking models. 
We perform a regression model to explain the REM metric (dependent variable)  using nine datasets characteristics  (independent variables). The latter are presented in Table \ref{tab:reg} and  include dataset-based measures inspired from \cite{Ado12,Yashar20} and effectiveness-based measures using the BM25 model.
  \begin{table}[!t]
 \centering
 \footnotesize
     \caption{Linear regression explaining catastrophic forgetting (REM metric) at the left dataset level. Significance: $***: p\leq 0.001;  **:0.001<p \leq 0.01; *: 0.01<p\leq 0.5$}
     \begin{adjustbox}{max width=\textwidth}
      \begin{tabular}{ll l l l H  } 
     \toprule
      &&\textbf{Characteristic} & \textbf{Description} &
      \textbf{Coeff} & \textbf{Coeff for REMN} \\
     \midrule
&&$R^{2}$ &&~0.544&~0.466 \\
\hline
\parbox[t]{2mm}{\multirow{10}{*}{\rotatebox[origin=c]{90}{Independent variables}}}&\parbox[t]{2mm}{\multirow{7}{*}{\rotatebox[origin=c]{90}{Dataset}}}&Constant & &~0.7014$***$  & ~0.6120$***$\\
&&RS& Retrieval space size: $log_{10}(D\times Q)$&-0.1883 & -0.3567$*$\\
&&RD&Relevance density: $log_{10}\frac{Qrels}{D\times Q}$&-0.3997$*$  &-0.0680\\
&&SD& Score relevance divergence: $KL(RSV_{D+},RSV_{D-})$&~0.0009 & ~0.0008\\
&&Vocab&Size of the vocabulary&-0.0932$*$ & -0.0527\\
&&DL&Average length of documents&-0.0349&-0.0814$*$\\
&&QL&Average length of queries&~0.1803$*$&~0.3143$*$\\
&&QD&Average query difficulty: $avg_q(\frac{1}{|q|}\sum_{w \in q} idf_w)$&~0.0044&-0.0187$*$\\
\cline{3-5}
&\parbox[t]{2mm}{\multirow{2}{*}{\rotatebox[origin=c]{90}{Eff.}}}&MAP&Effectiveness of the BM25: MAP&-0.0220$*$ &~0.0102\\
&&std-AP&Variation of BM25 effectiveness (AP metric): $\sigma_q(AP_q)$&~0.0543$*$ & ~0.0614$*$\\
\hline
\parbox[t]{2mm}{\multirow{7}{*}{\rotatebox[origin=c]{90}{Residual Variables}}}&&\multirow{2}{*}{$Dataset_i$} & MSmarco&~0.1803$8$ &~0.3143$*$\\
&&&Microblog&~0.5211$**$&~0.2977$*$  \\
\cline{3-5}
&&\multirow{5}{*}{$M_j$}&DRMM & ~0.1798$***$&~0.1445$***$\\
&&&PACRR & ~0.1965$***$&~0.1802$***$\\
&&&KNRM &~0.1924$***$&~0.1784$***$ \\
&&&VBERT & ~0.1313$***$&~0.0881$**$\\
&&&CEDR &~0.0014&~0.0207$*$ \\
\bottomrule
     \label{tab:reg}
     \end{tabular}
     \end{adjustbox}
     \vspace{-1cm}
 \end{table}
 To  artificially-generate datasets with varying data characteristics, we  follow the procedure detailed in  \cite{Ado12}: we sample queries within each left dataset in the settings presented in Table 1 (15 for $mb$ and 50 for $ms$) to create sub-datasets composed of those selected queries and the 100 top corresponding documents retrieved by the BM25 model. 
 
 Then, we replace in each setting the left dataset by the corresponding sub-dataset. We estimate for each model-setting pair the REM value as well as the characteristic values of the left sub-dataset. We repeat this procedure 300 times to obtain 300 new settings per model, based on the 300 sampled sub-datasets. This leads to  300 (sub-setting-model) pairs with a  variation for both the dependent  and the independent variables. 
 Finally, we build the following explanatory regression model, referring  to the "across dataset analysis" in \cite{Ado12}:
\vspace{-0.2cm}
\begin{equation}
     REM_{ij}= \sum_{k} C_k f_{ik} + Dataset_i + M_j + \epsilon_i
     \vspace{-0.3cm}
  \end{equation}
 where $i$ denotes the $i^{th}$ sub-setting  and   $j$ refers to the neural ranking model $M_j$. 
 $C_k$ and $f_{ik}$ denote respectively the weight  and the value of the $k^{th}$ characteristic of the left dataset in the $i^{th}$ sub-setting. Please note, that dataset feature values are independent of the model $M_j$.
  $Dataset_i$ and $M_j$ are the residual variables of resp. the left dataset and the model. The characteristic values $f_{ik}$ are centered before the regression as suggested in Adomavicius and Zhang \cite{Ado12}. 
  
  Table \ref{tab:reg} presents the result of the regression model. From $R^{2}$ and $Constant$, we can see that our regression model can explain $54.4\%$ of the variation of the REM metric, highlighting an overall good performance in explaining the remembering metric with  a good level of prediction ($0.7014$). 
  From the independent variables, we can infer that the difficulty of the dataset positively impacts the remembering (namely, decreasing the catastrophic forgetting). More precisely, lower the relevance density (RD), the BM25 effectiveness (MAP) and higher the variation in terms of BM25 performances over queries (std-AP) are, the higher the REM metric is. This suggests that relevance-matching difficulty provides positive feedback signals to the neural model  to  face diverse learning instances, and therefore  to better generalize over different application domains. This is however true to the constraint that the vocabulary of the dataset ($Vocab$) is not too large so as to boost neural ranking performance as outlined in \cite{Hofstatter-term,mitra-updated}. 
 Looking at the residual variables ($Dataset_j$ and $M_j$), we can corroborate the observations made at a first glance in RQ1 regarding the model families clearly opposing (DRMM-PACRR-KNRM-VBERT) and CEDR since the former statistically exhibit higher REM metrics values than CEDR. 

\vspace{-0.4cm}
\subsection{Mitigating Catastrophic Forgetting  (RQ3)}
\vspace{-0.2cm}
From RQ1, we observed that some models are more prone to the catastrophic forgetting problem than others. Our objective here  is to examine whether an EWC-based lifelong strategy can mitigate the problem. It is worth mentioning that this objective has been targeted in previous research in computer vision but without establishing a consensus \cite{Lee2017,thompson-etal-2019-overcoming,abs-1805-07441}. While some studies reveal that EWC outperforms domain adaptation strategies in their settings \cite{Lee2017,thompson-etal-2019-overcoming}, others found that it is less effective \cite{abs-1805-07441}. 
To achieve the experiment's objective, we particularly report the following measures in addition to the $REM/REMN$ measures: 1) $\Delta_{REM(REMN)}$  that reveals the improvement (positive or negative) of the $REM/REMN$ measures achieved using an EWC-based lifelong learning strategy over the $REM/REMN$ measure achieved using a fine-tuning strategy; 2) the PR measure introduced in Section 4.1. Unlikely, our aim through this measure here, is to highlight the performance stability of the learned model on the right dataset while avoiding catastrophic forgetting on the left dataset. 

We  turn now our attention to the \textbf{"EWC-based lifelong learning"} columns in Table \ref{tab:omega_results_map}.  
Our experiment results show that among the $9$ (resp. $11$) settings that exhibit  catastrophic forgetting in the combined model (resp. neural model), EWC strategy allows to improve $9/9$ i.e., $100\%$ (resp. $9/11$ i.e., $88\%$) of them in the range  $[+0.3\%, +96.1\%]$ (resp.$[+3.3\%, +79.7\%]$). Interestingly, this improvement in performance on the left dataset does not come at the cost of a significant decrease in performance on the right dataset since $100\%$ of the models achieve a $PR$ ratio greater than $0.96$. Given, in the one hand, the high variability  of the settings derived from the samples, and in the other hand, the very low number of settings ($10\%$ i.e., $2/20$) where a performance decrease is observed in the left dataset, we could argue that the EWC-based lifelong learning is not inherently impacted by dataset order leading to a general effectiveness trend over the models.  We emphasize this general trend by particularly  looking at the CEDR model which we recall (See Section 4.1, RQ1), clearly exhibits the catastrophic forgetting problem. As can be seen from Table \ref{tab:omega_results_map}, model performances on the left datasets are significantly improved ($+6.4\%\leq \Delta_{REM}\leq +96.1\%$; $0\%\leq \Delta_{REMN}\leq +8.7\%$ ) while keeping model performances on the right dataset stable ($0.961\leq PR \leq 1.008$).  This property is referred to as the stability-plasticity dilemma \cite{ParisiKPKW19}.

\begin{figure}[t]
\centering
\subfloat[Loss function]{ \includegraphics[width=0.43\linewidth]{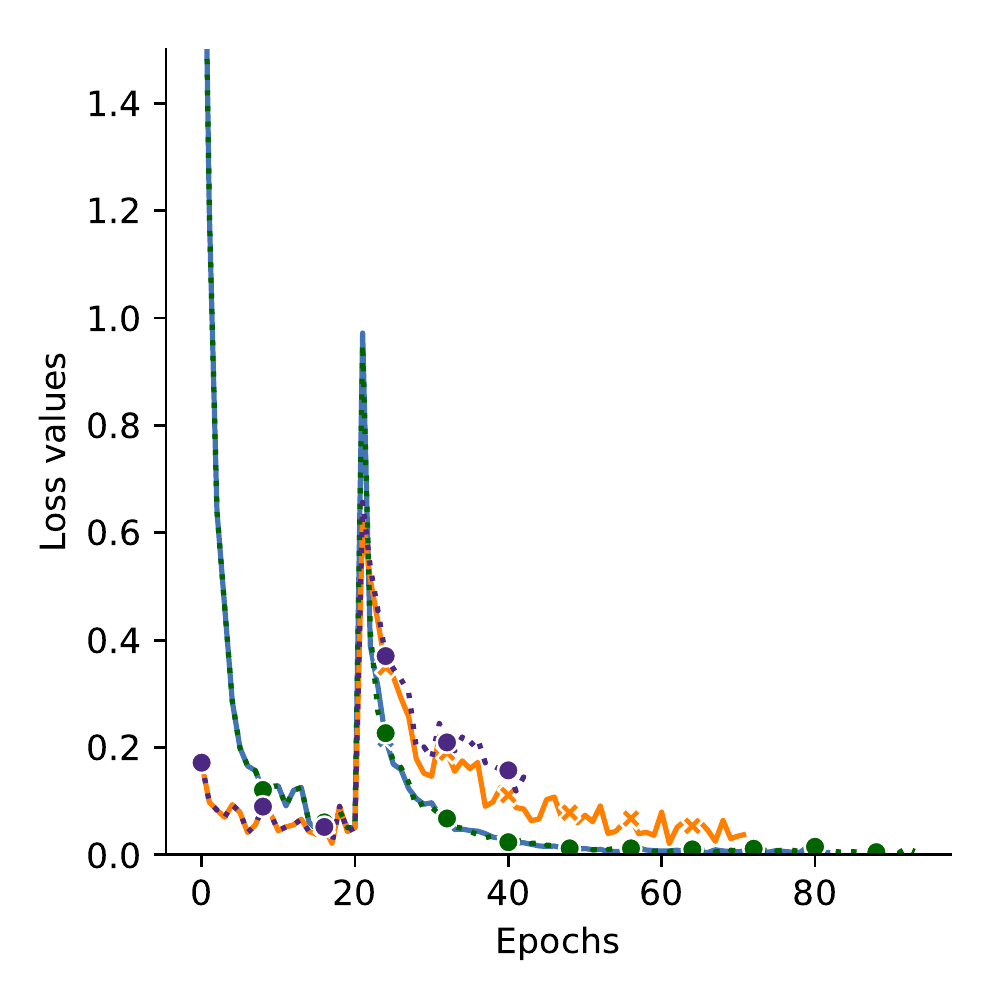}}
\subfloat[Performance on the $mb$ dataset]{ \includegraphics[width=0.5\linewidth]{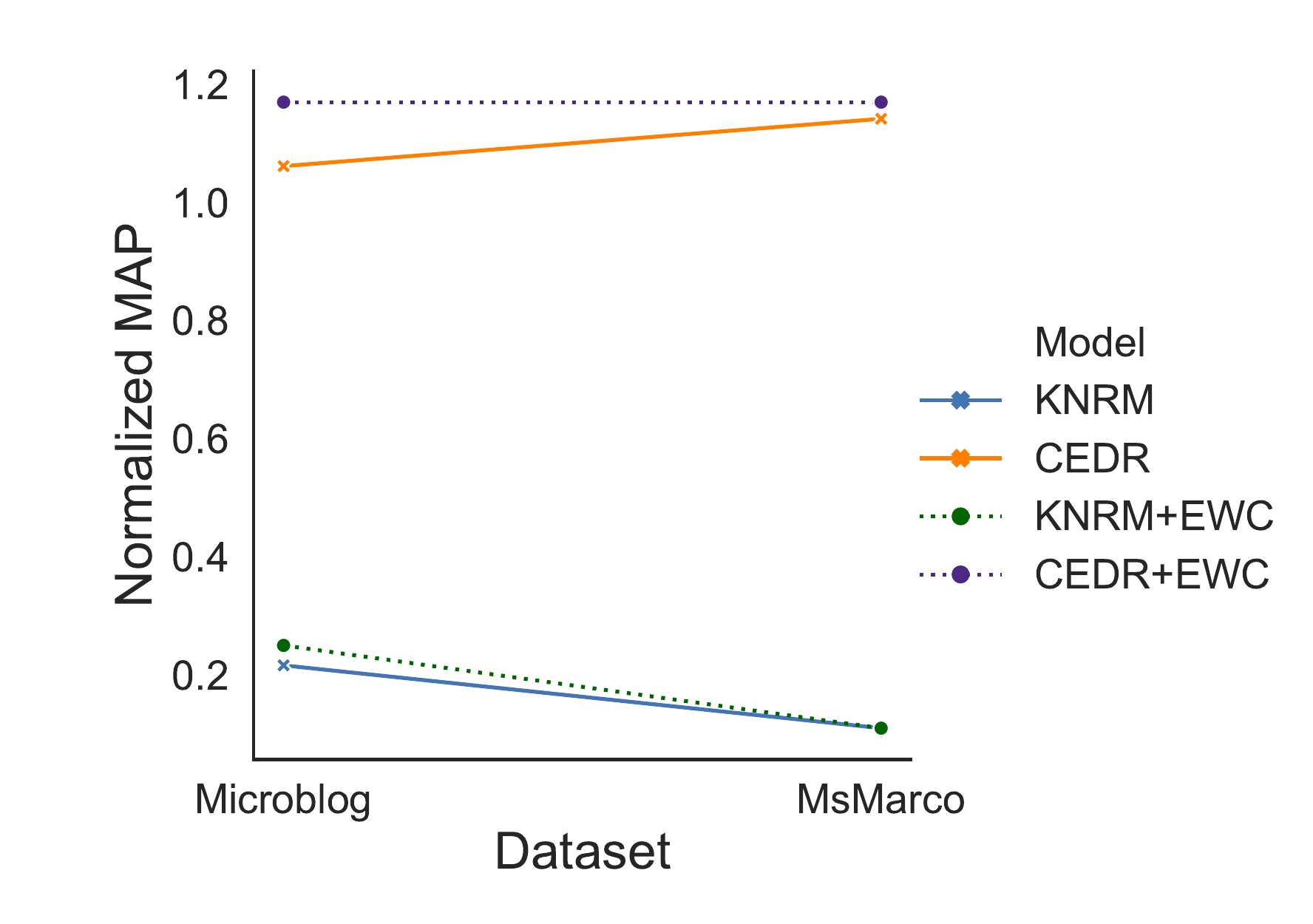}}
\vspace{-0.3cm}
\caption{Impact of the EWC strategy on loss and performance for the $ms \rightarrow mb$ setting.}
\label{fig:ewc-comparison}
\vspace{-0.6cm}
\end{figure}

To get a better overview of the effect of the EWC strategy, we compare in Figure \ref{fig:ewc-comparison} the behavior of the CEDR and KNRM models  which exhibit respectively low level  ($REM=0.510$) and high level of remembering ($REM=1$)  particularly in the setting $ms\rightarrow mb$. 
The loss curves in Figure 2(a) highlight a peak after the $20^{th}$ epoch for both CEDR and KNRM. This peak denotes the beginning of the fine-tuning on the $mb$ dataset. After this peak, we can observe that the curve representing the EWC-based CEDR loss (in purple)  is slightly above the CEDR loss (in orange), while both curves  for the KNRM model (green and blue resp. for with and without EWC) are overlayed. Combined with the statements outlined in RQ1 concerning the ability of the CEDR model to accumulate knowledge, this suggests that EWC is able to discriminate models prone to catastrophic forgetting and, when necessary, to relax the constraint of good ranking prediction on the dataset used for the fine-tuning to avoid over-fitting. 
This small degradation of knowledge acquisition during the fine-tuning  on the $ms$ dataset is carried out at the benefit of the previous knowledge retention to maintain retrieval performance on the $mb$ dataset  (Figure 2(b)). Thus, we can infer  that the   EWC strategy applied on neural ranking models plays fully its role to mitigate the trade-off between   stability and plasticity.

\vspace{-0.3cm}
\section{Conclusion}
\vspace{-0.2cm}
We investigated the problem of catastrophic forgetting in neural-network based ranking models. We carried out experiments using 5 SOTA  models and 3 datasets  showing that neural ranking effectiveness comes at the cost of forget and that transformer-based models allow a good balance between effectiveness and remembering. We also show that the EWC-based strategy mitigates the catastrophic forgetting problem while ensuring a good trade-off between transferability and plasticity. Besides, datasets providing weak and varying relevance signals are likely to be remembered. While previous work in the IR community mainly criticized neural models regarding effectiveness \cite{mitra2018,Onal2017NeuralIR,yang2019critically},  we provide complementary insights on the relationship between effectiveness and transferability  in a lifelong learning setting that involves cross-domain adaptation. 
We believe that our study, even under limited  setups, provides fair and generalizable results that could serve future research and system-design in neural IR. 

\vspace{-0.4cm}
\section{Acknowledgement}
\vspace{-0.4cm}
\begin{small}
We would like to thank projects ANR COST (ANR-18-CE23-0016) and ANR JCJC SESAMS (ANR-18- CE23-0001) for supporting this work.
\end{small}

\end{sloppypar}
\newpage
%
%
%
\bibliographystyle{splncs04.bst}
\bibliography{ECIR_2021_LL}

\end{document}